\documentclass[aps,twocolumn,prx,floatfix,showpacs,lengthcheck,citeautoscript,superscriptaddress,longbibliography]{revtex4-2}
\usepackage{scalerel, stackengine, amsmath}
\usepackage{amssymb}
\usepackage{physics}
\usepackage{graphicx}
\usepackage{bm}
\usepackage{color,soul}

\usepackage[dvipsnames,svgnames,table]{xcolor}
\usepackage{times}
\usepackage{MnSymbol}
\usepackage{bbold}
\usepackage[english]{babel}
\usepackage{hyperref}

\hypersetup{
pdfstartview={FitH},% fits the width of the page to the window
colorlinks=true,    % false: boxed links; true: colored links
linkcolor=NavyBlue, % color of internal links
citecolor=NavyBlue,   % color of links to bibliography
filecolor=NavyBlue, % color of file links
urlcolor=NavyBlue   % color of external links
}

\newcommand{\smeq}{\! = \!}

\newcommand{\smpl}{\! + \!}
\newcommand{\smmi}{\! - \!}

\newcommand{\ve}{\varepsilon}

\newcommand{\bea}{\begin{eqnarray}}
\newcommand{\eea}{\end{eqnarray}}
\newcommand{\be}{\begin{equation}}
\newcommand{\ee}{\end{equation}}

\newcommand{\Z}{\mathbb{Z}}
\newcommand{\U}{\mathrm{U}}

\newcommand{\heff}{\hat H_{\mathrm{eff}}}
\newcommand{\vp}{\varphi}
\newcommand{\h}{\hat}
\newcommand{\hata}{\hat{a}}

\newcommand{\hatb}{\hat{b}}
\newcommand{\dg}{\dagger}
\newcommand{\hc}{\mbox{h.c.}}

\definecolor{Nathanblue}{rgb}{0.094,0.317,0.5607}
\newcommand{\blue}{\color{Nathanblue}}

\begin{document}

\title{{\blue Dynamic Realization of Majorana Zero Modes in a Particle-Conserving Ladder}}

\author{Anaïs Defossez}
\affiliation{Laboratoire Kastler Brossel, Coll\`ege de France, CNRS, ENS-Universit\'e PSL,
Sorbonne Universit\'e, 11 Place Marcelin Berthelot, 75005 Paris, France
}
\affiliation{International Solvay Institutes, 1050 Brussels, Belgium}
\affiliation{Center for Nonlinear Phenomena and Complex Systems, Universit\'e Libre de Bruxelles, CP 231, Campus Plaine, B-1050 Brussels, Belgium}
\author{Laurens Vanderstraeten}
\affiliation{Center for Nonlinear Phenomena and Complex Systems, Universit\'e Libre de Bruxelles, CP 231, Campus Plaine, B-1050 Brussels, Belgium}
%\author{Leonardo Mazza}
\author{Lucila {Peralta Gavensky}}
\affiliation{International Solvay Institutes, 1050 Brussels, Belgium}
\affiliation{Center for Nonlinear Phenomena and Complex Systems, Universit\'e Libre de Bruxelles, CP 231, Campus Plaine, B-1050 Brussels, Belgium}
\author{Nathan Goldman}
\email{nathan.goldman@lkb.ens.be}
\affiliation{Laboratoire Kastler Brossel, Coll\`ege de France, CNRS, ENS-Universit\'e PSL,
Sorbonne Universit\'e, 11 Place Marcelin Berthelot, 75005 Paris, France
}
\affiliation{International Solvay Institutes, 1050 Brussels, Belgium}
\affiliation{Center for Nonlinear Phenomena and Complex Systems, Universit\'e Libre de Bruxelles, CP 231, Campus Plaine, B-1050 Brussels, Belgium}

\begin{abstract}
We present a scheme to realize a topological superconducting system supporting Majorana zero modes, within a number-conserving framework suitable for optical-lattice experiments. Our approach builds on the engineering of pair-hopping processes on a ladder geometry, using a sequence of pulses that activate single-particle hopping in a time-periodic manner. We demonstrate that this dynamic setting is well captured by an effective Hamiltonian that preserves the parity symmetry, a key requirement for the stabilization of Majorana zero modes. The phase diagram of our system is determined using a bosonization theory, which is then validated by a numerical study of the topological bulk gap and entanglement spectrum using matrix product states. Our results indicate that Majorana zero modes can be stabilized in a large parameter space, accessible in optical-lattice experiments. 
\end{abstract}
\maketitle

\section{Introduction}

The search for Majorana zero modes (MZMs) in condensed-matter systems remains a central challenge in modern physics. These charge-neutral quasiparticles naturally arise as boundary excitations of topological superconductors~\cite{Kitaev2001, Leijnse2012, Beenakker2013, Sato2017}, and they have drawn considerable attention due to their non-Abelian braiding properties, which have potential applications in topological quantum computation~\cite{Das_Sarma_2008, Alicea2011, Kraus_2013_braiding, Mazza_2016}.
Building on Kitaev's seminal work~\cite{Kitaev2001}, most approaches to realizing MZMs rely on a mean-field picture of the superconducting state: from proximitized semiconducting nanowires in solid-state devices~\cite{Oreg2010,Lutchyn2010} to one-dimensional optical lattices coupled to BCS or molecular atomic reservoirs~\cite{Jiang2011,Kraus_2012,Nascimbene_2013}. In this grand-canonical framework, the U(1) symmetry associated to particle-number conservation is reduced to a $\mathbb{Z}_2$ symmetry reflecting the  conservation of total fermion parity.

Beyond these mean-field approaches, a growing body of theoretical work has demonstrated the emergence of topologically-ordered states supporting MZMs in interacting, number-conserving fermionic systems, offering an interesting route for their realization in cold-atom platforms. In this many-body context, the existence of Majorana edge modes has been established through various methods, including bosonization techniques ~\cite{Cheng2011, Fidkowski2011, Sau2011,Kraus2013,liu2019phase,Tausendpfund_2023}, numerical simulations~\cite{Kraus2013, Iemini2017, Lisandrini2022} and the construction of exactly solvable models~\citep{Ortiz2014, Ortiz2016, Diehl_2015, Lang2015}. A particularly simple and minimal setting was put forward in Ref.~\citep{Cheng2011}, where spinless fermions are defined on a two-leg ladder, with intra-leg hopping and local inter-leg pair-hopping processes. Importantly, this setting preserves fermion parity within each leg, a requirement for topological superconductivity with stable MZMs.  Theoretical works proposed schemes to realize such parity-preserving fermionic ladders in an approximate manner, by suppressing inter-chain single-particle hopping through energetic constraints~\cite{Kraus2013} or Aharonov-Bohm caging~\cite{Tausendpfund_2023}.

\begin{figure}[h!]
    \centering
    \includegraphics[scale=0.45]{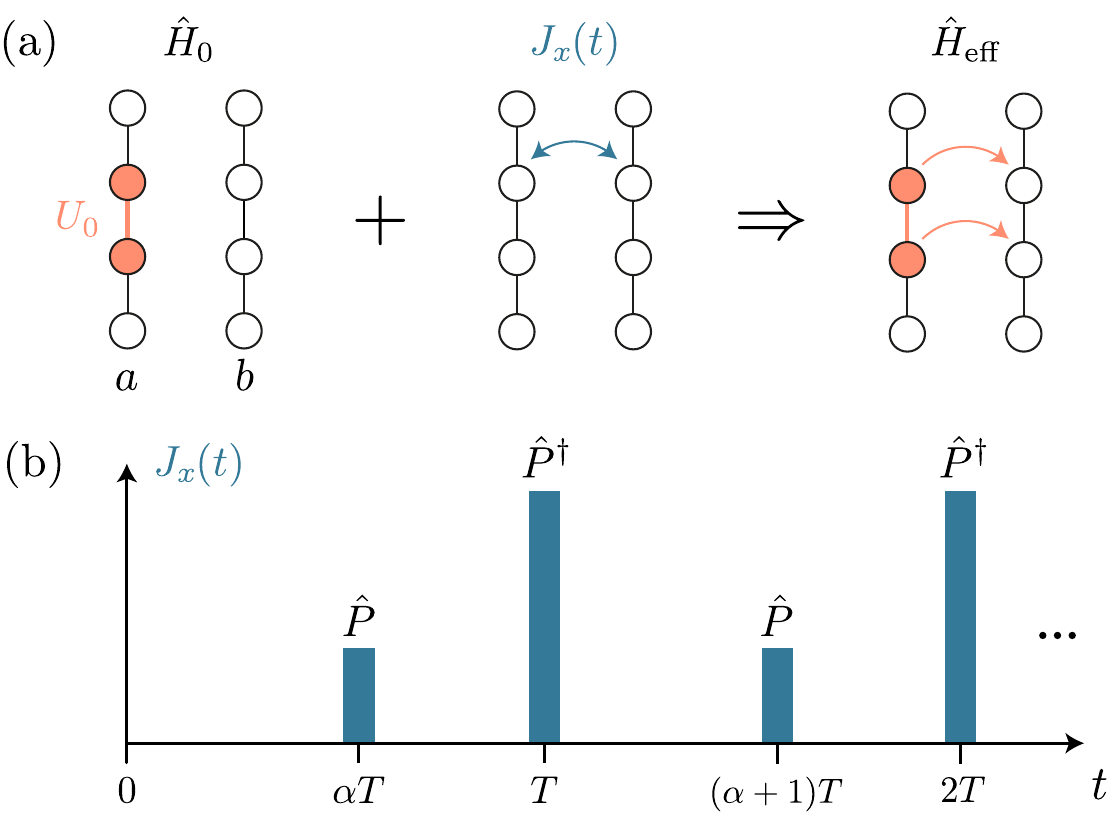}
    \caption{(a) Sketch of the dynamic approach. The bare Hamiltonian $\hat H_0$ describes two decoupled fermionic chains, exhibiting nearest-neighbor interactions of strength $U_0$; activating inter-chain hopping $J_x(t)$, in a time-periodic manner, allows to generate effective pair-hopping processes, eventually leading to a topological superconducting phase. (b) The time-periodic sequence describing the activation of the inter-chain hopping $(\hat P^{(\dagger)} )$ is characterized by the period $T$ and the drive parameter $\alpha$.}
    \label{fig_sketch}
\end{figure}

In this Letter, we propose an alternative strategy to realize a number-conserving ladder system featuring MZMs, using the tools of Floquet engineering. Subjecting a physical system to a time-periodic drive has been widely explored to engineer exotic band structures and interaction processes in various physical contexts, ranging from ultracold atoms~\cite{Eckardt_review,Cooper_review,weitenberg2021tailoring} to the solid state~\cite{Cayssol_2013,rudner2020band} and photonics~\cite{Ozawa_review}. Inspired by a prior work~\cite{Goldman_PRXQuantum}, we propose to generate pair-hopping processes on a lattice by combining bare density-density interactions with a sequence of pulsed single-particle hopping processes. We demonstrate that such a scheme can be applied to a fermionic ladder in view of realizing a parity-preserving system hosting stable MZMs. We discuss the role of the micromotion in this time-periodic framework, demonstrating parity-preserving dynamics at stroboscopic observation times. We then explore the phase diagram of this model using a bosonization approach, identifying a topological superconducting phase in a large parameter space. These predictions are then confirmed through tensor-network studies of the topological gap, entanglement spectrum and edge-mode correlations.

\section{The model and effective Hamiltonian} 

We start by considering two decoupled fermionic wires (denoted $a$ and $b$), containing $L$ sites each, and described by the following number-conserving Hamiltonian
\begin{align}
    \h H_0 &= -\tau \sum_j \left( \h a^\dg_j \h a_{j+1} + \h b^\dg_j \h b_{j+1} + \hc \right) \notag \\ 
    &\, + U_0 \sum_j \left( \h a^\dg_j \h a^\dg_{j+1} \h a_{j+1} \h a_j + \h b^\dg_j \h b^\dg_{j+1} \h b_{j+1} \h b_j\right)\, , 
    \label{eq: H_0}
\end{align}
where $\hat a^{(\dg)}_j$ and $\hat b^{(\dg)}_j$ annihilate (or create) a fermion at the lattice site $j$, on the wires $a$ and $b$, respectively. These fermionic operators satisfy the usual anti-commutation relations, e.g.~$\left\{ \h a^\dg_j , \h a_k \right\} \!=\! \delta_{j,k}$.  The first line in Eq.~\eqref{eq: H_0} describes intra-chain single-particle hopping with amplitude $\tau$; the second line describes intra-chain nearest-neighbor interactions of strength $U_0$.

For the sake of later convenience, we introduce a set of spin operators \cite{Auerbach_1994}
\begin{align}
    \h J_x^j &= \frac{1}{2} \left( \h a^\dg_j \h b_j + \h b^\dg_j \h a_j \right) \quad ; \quad \h J_y^j = -\frac{i}{2} \left( \h a^\dg_j \h b_j - \h b^\dg_j \h a_j \right) \notag \\ 
    \h J_z^j &= \frac{1}{2} \left( \h a^\dg_j \h a_j - \h b^\dg_j \h b_j \right) \quad ; \quad \h N^j = \h a^\dg_j \h a_j + \h b^\dg_j \h b_j \, .
    \label{eq: spin op.}
\end{align}
It is readily verified that these operators satisfy the commutation relations $[ \h J_\mu^j, \h J_\nu^k ] \!=\! i \delta^{j,k} \ve_{\mu \nu \rho} \h J_\rho^k$. While the intra-chain single-particle hopping does not have a spin representation, the intra-chain interactions in Eq.~\eqref{eq: H_0} can be written in the simple form $U_0/2 \sum_j ( 4 \h J_z^j \h J_z^{j+1} + \h N^j \h N^{j+1} )$. 

The main target of this work concerns the realization of a number-conserving topological ladder system~\cite{Kraus2013}, starting from the two decoupled fermionic wires in Eq.~\eqref{eq: H_0}. The key ingredient that is responsible for the existence and stability of topological superconducting phases, and which requires some engineering, are the pair-hopping processes connecting the two wires~\cite{Cheng2011, Kraus2013}. Inspired by Ref.~\cite{Goldman_PRXQuantum}, we propose to generate the required pair-hopping processes through a time-periodic sequence, which activates single-particle hopping processes in a fast and pulsed manner; see Fig.~\ref{fig_sketch}(a). We introduce a hierarchy of time scales, $t_p \! \ll \! T\! \ll \!t_{\rm{ch}}$, where $T$ denotes the driving period, $t_p$ is the pulse duration and $t_{\rm{ch}}$ is a characteristic time scale (e.g.~the hopping time $1/\tau$). The pulsed coupling operator is chosen in the form $\h P \! \equiv \! e^{i \eta \h J_x}$, where $\h J_x \!\equiv \! \sum_j \h J_x^j$ denotes the inter-leg hopping terms, and where $\eta \in \mathbb R$ is a drive parameter controlled by the strength and duration of the pulsed hopping.

Concretely, the time-evolution operator over one driving period is designed according to the sequence [Fig.~\ref{fig_sketch}(b)]
\begin{align}
    \h U(T) &= \h P^\dg \, e^{-i(1-\alpha)T \h H_0} \, \h P \, e^{-i \alpha T \h H_0} \label{eq: U(T) pulse} \\ 
    &= e^{-i (1-\alpha)T \h H_1} e^{-i \alpha T \h H_0} \, ,\notag
\end{align}
where $\h H_1 \! \equiv \! \h P^\dg \h H_0 \h P$, and where $\alpha \in [0,1]$ denotes a second drive parameter; see Fig.~\ref{fig_sketch}(b). We note that $\h P^\dg \!=\! e^{-i \eta \h J_x}$ is equivalently given by $e^{i(-\eta + 2\pi m) \h J_x}$, $m \in \mathbb Z$, such that the activated-hopping amplitude does not need to change sign over the duration of the sequence, which is experimentally convenient; see Fig.~\ref{fig_sketch}(b) and Ref.~\cite{Goldman_PRXQuantum}. In Eq.~\eqref{eq: U(T) pulse}, the pulses are assumed to be instantaneous; the effects associated with a finite pulse duration $t_p$ are discussed in Appendix~\ref{app:pulse}.

In the high-frequency regime of the drive, one can apply the Trotter approximation to Eq.~\eqref{eq: U(T) pulse} in view of deriving an effective (Floquet) Hamiltonian $U(T) \!=\! e^{-iT \h H_{\rm{eff}}}$, which is then simply given by $\h H_{\rm{eff}} \!=\! \alpha \h H_0 + (1-\alpha) \h H_1$, where we have neglected corrections of order $O(T)$. Since $\h J_x$ commutes with $\h N^j$ and with the intra-chain single-particle hopping terms, the computation of $\h H_1$ can be directly obtained from the relation 
\begin{align}
    e&^{-i \eta \h J_x} \h J_z^j \h J_z^{j+1} e^{i \eta \h J_x} = \frac{1}{2} \cos 2\eta \left( \h J_z^j \h J_z^{j+1} - \h J_y^j \h J_y^{j+1} \right)  \label{eq: BCH} \\ 
    &- \frac{1}{2} \sin 2 \eta \left( \h J_y^j \h J_z^{j+1} + \h J_z^j \h J_y^{j+1} \right) + \frac{1}{2} \left( \h J_z^j \h J_z^{j+1} + \h J_y^j \h J_y^{j+1} \right) \notag.
\end{align}
which derives from the Baker-Campbell-Hausdorff formula \cite{Sakurai_Napolitano_2017}. 

Importantly, the effective interactions in Eq.~\eqref{eq: BCH} exhibit $\mathbb Z_2$-breaking processes, which are all contained in the terms proportional to $\sin(2\eta)$. We remind that the $\mathbb Z_2$ symmetry, which is associated with the leg-parity operator $\h{\mathcal P} \smeq (-1)^{\hat N_a} \smeq \pm (-1)^{\hat N_b}$, is crucial to preserve the ground-state degeneracy associated with the presence of MZMs~\cite{Cheng2011}. These undesired $\mathbb Z_2$-breaking processes can thus be annihilated by simply tuning the drive parameter to the value $\eta \smeq \pi/2$.

In the original fermionic representation, the resulting effective Hamiltonian thus reads 
\begin{align}
    &\h H_{\rm{eff}} = -\tau \sum_j \left( \h a^{\dg}_j \h a_{j+1} + b^{\dg}_j \h b_{j+1} + \hc \right) \label{eq: H_eff ferm. rep.} \\ 
    &\, + U_1 \sum_j \left( \h n^a_j \h n^a_{j+1} + \h n^b_j \h n^b_{j+1} \right) + U_2 \sum_j \left( \h n^a_j \h n^b_{j+1} + \h n^b_j \h n^a_{j+1} \right) \notag \\ 
    &\, + U_2 \sum_j \left( \h a^{\dg}_j \h b^{\dg}_{j+1} \h a_{j+1} \h b_j - \h a^{\dg}_j \h a^{\dg}_{j+1} \h b_{j+1} \h b_j + \hc \right) \, ,\notag   
\end{align}
where $U_1 \!=\! \frac{U_0}{2}(1+ \alpha)$ and $U_2 \!=\! \frac{U_0}{2}(1 - \alpha)$, and $\h n^\beta \!=\! \h \beta^\dg_j \h \beta_j$ is a fermionic density operator on the wire $\beta \smeq a,b$. The effective Hamiltonian established in Eq.~\eqref{eq: H_eff ferm. rep.} displays novel two-body processes, including inter-chain density-density interactions, inter-chain swapping processes, and most importantly, the desired pair-hopping processes. We emphasize that all these interaction processes couple the two wires in a number- and parity-conserving manner. We also point out that the effective Hamiltonian is trivial in the undriven cases $\alpha\!=\!0,1$, and that $\h H_{\rm{eff}} (\alpha)$ is equivalent to $\h H_{\rm{eff}} (1-\alpha)$ up to a unitary transformation.

Crucially, while Eq.~\eqref{eq: H_eff ferm. rep.} was obtained using the Trotter approximation, we find that the effective Hamiltonian $\h H_{\rm{eff}}$ preserves the $\mathbb Z_2$ (parity) symmetry at all orders of the high-frequency expansion~\cite{Goldman_Dalibard_PRX}. This can be appreciated by noting that if $\h O_1$ and $\h O_2$ are two parity-conserving operators, then so is their commutator $[\h O_1, \h O_2]$.

%%%%%%%%%%%%%%%%%%%%%%%%%%%%%%%%

We point out that the periodic-driving scheme discussed above could be modified and generalized in various ways. For instance, combining the pulse sequence in Eq.~\eqref{eq: U(T) pulse} with an additional time modulation of the bare interaction strength $U_0 (t)$ would offer individual control over the different interaction processes displayed in Eq.~\eqref{eq: H_eff ferm. rep.}; see Appendix~\ref{app_pair}. Such a control over inter-site interactions would be possible in ultracold dipolar gases trapped in optical lattices~\cite{Su_2023}, where such processes rely on the orientation of the dipoles, and can thus be manipulated. Besides, we note that our pulse sequence could be substituted by a  continuous time-periodic drive; see Appendix~\ref{app_cont}. In this case, the model preserves $\mathbb Z_2$ symmetry at lowest-order in the high-frequency expansion, without relying on any fine-tuned parameter.

%%%%%%%%%%%%%%%%%%%%%%%%%%%%%%%%

\section{Validity of the effective model and preservation of $\mathbb Z_2$ symmetry} 

We now numerically explore the accuracy of the effective Hamiltonian in Eq.~\eqref{eq: H_eff ferm. rep.}, by studying the time evolution of two interacting fermions on a square plaquette (2 sites per chain). The main focus is set on validating the preservation of $\mathbb Z_2$-symmetry at stroboscopic times throughout the time evolution, when setting the drive parameter close to $\eta \!=\! \pi/2$, and the role of micromotion. 

\begin{figure}[t]
    \centering
    \includegraphics[width=0.9999\columnwidth]{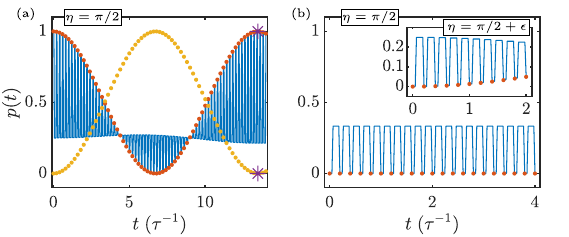}
    \caption{(a) Rabi oscillations between the states $\h a^\dg_1 \h a^\dg_2 \ket{0}$ and $\h b^\dg_1 \h b^\dg_2 \ket{0}$, when setting $\eta \!=\! \pi/2$. The Rabi period $T_R\!=\! 2 \pi / |(U_0 (1-\alpha))|$ is indicated by the purple stars. (b) Time evolution of the mean probability of changing parity for $\eta \!=\! \pi/2$, and for $\eta \!=\! \pi/2 + 0.1$ (inset). In both panels, we compare  the exact dynamics generated by the full time-dependent Hamiltonian (blue curve),  including the micromotion-, with the stroboscopic dynamics ($t_n\!=\!nT$, $n \in \mathbb Z$) associated with the effective Hamiltonian in Eq.~\eqref{eq: H_eff ferm. rep.} (red and yellow dots). The driving parameters are $T\!=\!0.2$ and $\alpha \!=\! 1/3$, and we set $U_0\!=\!-0.7$ and $\tau \!=\! 1$.}
    \label{fig: validity}
\end{figure}

Figure~\ref{fig: validity}(a) illustrates the Rabi oscillations between the two states $\h a^\dg_1 \h a^\dg_2 \ket{0}$ and $\h b^\dg_1 \h b^\dg_2 \ket{0}$, under the ideal condition $\eta \!=\! \pi/2$, and when probing the dynamics at stroboscopic times. The full-time dynamics also reveal the micromotion, which activates single-particle inter-chain hopping processes within each period of the drive. This result confirms the realization of effective pair-hopping processes and the decoupling of the two different parity sectors at stroboscopic times, thereby realizing $\mathbb Z_2$-preserving dynamics  over long evolution times in this optimal configuration. The numerical analysis further shows that the exact dynamics is well described by the effective Hamiltonian at stroboscopic times, deep in the high-frequency regime.

Then, we analyze in Fig.~\ref{fig: validity}(b) the probability of changing parity during the time evolution, demonstrating that the parity is indeed conserved at stroboscopic times when setting $\eta \!=\! \pi/2$. In practice, we average this probability over all possible initial states of our two-fermion setting. We point out that the dynamics breaks parity within each period of the drive, a direct manifestation of the single-particle inter-chain hopping generated by the micromotion. The inset compares the dynamics of this same observable for $\eta\!=\!\pi/2 + \epsilon$, with $\epsilon\!=\!0.1,$ highlighting the deviation from the ideal $\mathbb Z_2$-preserving dynamics.  In particular, this result indicates that $\mathbb Z_2$-preserving dynamics are observed over a time scale $\lesssim 1/\epsilon$.
 
In the following, we aim at identifying the system parameters for which the driven system enters a topological phase exhibiting MZMs. We will first gain intuition from a field-theoretical method based on bosonization, and then provide more concrete topological signatures using numerical tensor network methods. We point out that the studies presented in the following Sections~\ref{sect:boson}-\ref{sect:mps} are based on the low-energy properties of the effective Hamiltonian in Eq.~\eqref{eq: H_eff ferm. rep.}. In practice, this low-energy physics could be reached through adiabatic quantum state preparation, as we further discuss in Section~\ref{sect:conclusions}.

\section{Bosonization}\label{sect:boson} 

In this section, we use a low-energy description to qualitatively examine how the different interaction processes entering the effective Hamiltonian in Eq.~\eqref{eq: H_eff ferm. rep.} compete in view of forming a topological phase.  Following a standard bosonization approach~\cite{Giamarchi_2003}, the individual wires are described by the bosonic fields $\hat\vp_\beta$ and their dual $\hat\vartheta_\beta$. The inter-chain couplings are then introduced by going to the bonding and anti-bonding basis, 
\begin{equation}
    \hat\varphi_{\pm}=\frac{1}{\sqrt{2}}\left( \hat\varphi_a \pm \hat\varphi_b \right), \qquad \hat\vartheta_{\pm}=\frac{1}{\sqrt{2}}\left( \hat\vartheta_a \pm \hat\vartheta_b \right),
\end{equation}
yielding a total Hamiltonian density 
\begin{multline} \label{eq: bosonised eff Ham}
    \hat{\mathcal H}_{\rm{bos}} = \sum_{r = \pm} \frac{v_r}{2} \left[ K_r \left( \partial_x \hat\vartheta_r \right)^2 + \frac{1}{K_r} \left( \partial_x \hat\vp_r \right)^2 \right] \\ 
     - \frac{g_p}{2\pi^2} \cos \left( \sqrt{8\pi} \hat\vartheta_- \right) + \frac{g_{bs}}{2\pi^2} \cos \left( \sqrt{8\pi} \hat\vp_- \right)\, ,
\end{multline}
with the effective parameters
\begin{align} \label{eq: luttinger param}
    & \pi v_\pm K_\pm =v_F \pi + U_1 \cos(2ak_F) \mp \, 2U_2 \sin^2 ak_F  \\ 
    & \frac{\pi v_\pm}{K_\pm} = v_F \pi + U_1(2-\cos(2ak_F)) \pm 2U_2 \sin^2ak_F ,
\end{align}
and the couplings
\begin{equation}
    g_p = -4 U_2 \sin^2 ak_F \quad ; \quad g_{bs} = -4U_2 \cos^2 ak_F \, .
    \label{eq: bare param}
\end{equation}
The Fermi momentum $k_F$ is related to the filling through $a k_F\!\smeq\!\frac{a N\pi}{2L}\!\smeq\!\pi \nu$, where $a\!\rightarrow\! 0$ denotes the lattice spacing~\cite{Giamarchi_2003,Tausendpfund_2023}. In the bonding sector, the inter-chain processes simply lead to a rescaling of the effective parameters. In the anti-bonding sector, however, there are two sine-Gordon terms that can potentially open a gap. As shown in Refs.~\cite{Cheng2011,Kraus2013}, the pair-tunneling term with bare coupling $g_p$ drives the anti-bonding sector into a regime that is well described by the continuum limit of a Kitaev chain; this term thus generates a topological superconducting phase, exhibiting a spectral gap and Majorana edge modes. In contrast, the backscattering term with bare coupling $g_{bs}$ drives the system into a topologically trivial phase. 

In order to understand which of the two terms dominates, we derive the perturbative renormalization group (RG) equations. At lowest order in the couplings, and considering $K_- \! \approx \! 1$ (i.e.~the marginal regime), the RG flows are governed by the following equations \cite{Giamarchi_2003}
\begin{align}
    \begin{cases}
        \frac{dy_-}{dl} = 2 \left( y_p^2 - y_{bs}^2 \right) \\ 
        \frac{dy_p}{dl} = y_- y_p \\ 
        \frac{dy_{bs}}{dl} = -y_- y_{bs}
    \end{cases}\label{RG_eq}
\end{align}
where $l$ is the parameter that scales the cutoff in position space, $K_- \!=\! 1+\frac{y_-}{2}$, $y_p \!=\! \frac{g_p}{\pi v_-}$ and $y_{bs} \!=\! \frac{g_{bs}}{\pi v_-}$. From these equations, one readily obtains that $K_-\!>\!1$ is required to drive the system into the topological phase.

\begin{figure}[t]
    \centering
    \includegraphics[width=0.9999\columnwidth]{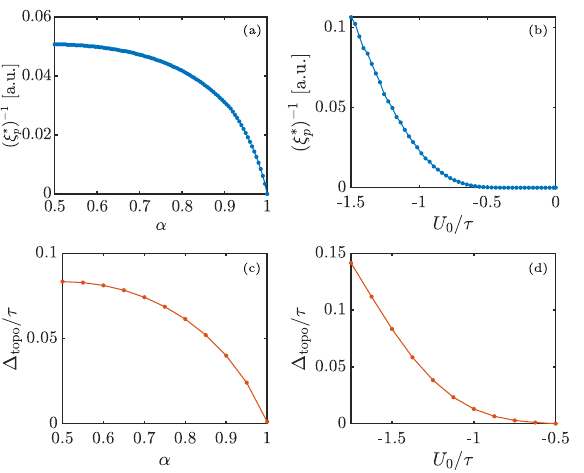}
    \caption{(a)-(b) Bosonization estimation of the inverse correlation length $(\xi^*_p)^{-1}$ associated to the topological gap:~(a) as a function of $\alpha \in [0.5,1]$ for attractive bare interactions $U_0\!=\!-1.2$; (b) as a function of $U_0 \in [-1.5,0]$ for $\alpha\!=\! 1/2$. The critical value is chosen to be $y_p(l^*) = 9$. (c)-(d) Infinite MPS simulations without imposing $\mathcal{P}$ symmetry:~(c) the topological gap as a function of $\alpha$ for attractive bare interactions $U_0\!=\!-1.5$; (d) the topological gap as a function of $U_0$ for fixed $\alpha\!=\!1/2$. All gaps are expressed in units of the hopping $\tau$. The filling is fixed to $\nu \smeq 1/3$ in all panels.}
    \label{fig:phase}
\end{figure}

One can obtain a qualitative phase diagram by monitoring the conditions under which one of the two coupling constants, $g_{bs}$ or $g_{p}$, flows to strong coupling. This is established by integrating Eqs.~\eqref{RG_eq}, using the bare values as initial conditions. Here, the strong-coupling regime is assumed to be reached at a critical value $y_{p,bs}(l^*) \! \gtrsim \! O(1)$, such that the perturbative RG equations remain reasonably valid. The resulting critical value $l^*$ yields a correlation length $\xi_{p,bc}^* \! \sim \! e^{l*}$, which scales as the inverse of the associated spectral gap~\cite{Malard_2013}. Figures~\ref{fig:phase}(a)-(b) show the behavior of the inverse correlation length $(\xi^*_p)^{-1}$ associated with the pair-hopping coupling, as a function of the parameters $\alpha$ and $U_0$, at fixed filling $\nu=1/3$. This quantity is related to the topological superconducting gap, $(\xi^*_p)^{-1}\! \sim\!\Delta_{{\rm topo}}$, which can be explicitly computed using tensor-network methods; see Figs.~\ref{fig:phase}(c)-(d) and below. We point out that the range of  interaction strength $U_0$ was chosen such that the approximation $K_- \! \approx \! 1 $ holds. The results in Fig.~\ref{fig:phase} suggest that the gap opens up  slowly and in a non-analytic manner for $U_0\!<\! 0$, and behaves as a power law $(1-\alpha^\kappa)$, with a maximal value at $\alpha\!\smeq\! 1/2$. In particular, we have verified that the transition between the gapped and the gapless regimes strictly occurs at $U_0\!=\!0$,  $\forall \alpha \neq 0,1$. This analysis points towards the fact that, similar to BCS superconductors,  bare attractive interactions $U_0\!<\!0$ promote the flow to a topological superconducting phase.

\section{Numerical studies and phase diagram}\label{sect:mps} %

We now confirm these field-theoretical predictions through numerical simulations, using tensor networks. Earlier numerical studies \cite{Kraus2013, Iemini2017, Lisandrini2022, Tausendpfund_2023} have used matrix product states (MPS) methods to find the lowest-lying states on a finite ladder system with open boundary conditions, analyzing the ground state degeneracy and entanglement spectrum as signatures of topological order. Here, we consider a ladder system directly in the thermodynamic limit, for which accurate variational ground-state approximations can be obtained in terms of infinite MPS \cite{ZaunerStauber2018, Vanderstraeten2019}. In the thermodynamic limit, the physics of the topological order is characterized by a spontaneous breaking of the parity symmetry in the ground state, yet without any local order parameter \cite{Ortiz2014}. In the topological phase $U_0\!<\!0$, we indeed find that the infinite MPS breaks the symmetry for any value of $\alpha$ in the range $0\!<\!\alpha\!<\!1$. In addition, we find that the local order parameter $\hat O_j\!=\! \hat b_j^\dagger \hat a_j$ \cite{Cheng2011} always yields a vanishing expectation value in the MPS ground states; see Appendix~\ref{app_MPS}.

\par As shown by the bosonization approach, the topological phase exhibits a gapless mode with central charge $c\!=\!1$, on top of which the gapped topological sector lies. The topological gap can be found in the single-particle sector of the bulk spectrum, and is defined as~\cite{Kraus2013}
\begin{equation}
    \Delta_{\mathrm{topo}}\!=\!(\Delta_{Q=+1}+\Delta_{Q=-1})/2,
\end{equation}
where $\Delta_Q$ denotes the bulk gap in the charge sector $Q$ above a given ground-state filling. On a finite system with open boundary conditions, the numerical extraction of the single-particle bulk gap is complicated due to the presence of the topological  ground-state degeneracy and a tower of excitations due to the gapless mode \cite{Kraus2013}. In the infinite system, however, one can target the lowest-lying bulk excitations on top of the symmetry-broken MPS ground state~\cite{Haegeman2012, ZaunerStauber2018b, Vanderstraeten2019}, and estimate the gaps in the different charge sectors directly. Following this approach, we have calculated the topological gap as a function of $\alpha$, and show the results in Fig.~\ref{fig:phase}(c) for fixed interaction strength $U_0\!=\!-1.5$ and filling $\nu\!=\!1/3$. We observe a power-law behavior close to $\alpha\!=\!1$, in agreement with the bosonization prediction, and a maximum at $\alpha\!=\!0.5$ with a sizable magnitude of $\Delta_{\rm topo}\!\approx\!0.1\tau$. In the neutral sector, we find that the gap $\Delta_{Q=0}$ nicely converges to zero for all values of $\alpha$ (not shown). In Fig.~\ref{fig:phase}(d), we present the gap as a function of $U_0$, showing a qualitative agreement with the bosonization prediction. Note that one would expect the topological phase to terminate around $U_0\!=\!-2$, the value for which the decoupled wires are known to exhibit phase separation \cite{Gotta2021}. In the regime around $U_0\!=\!-2$, the infinite-MPS simulations become unstable because of the proximity to this phase-separated region.

\par
\begin{figure}[t]
    \centering
    \includegraphics[width=0.999\columnwidth]{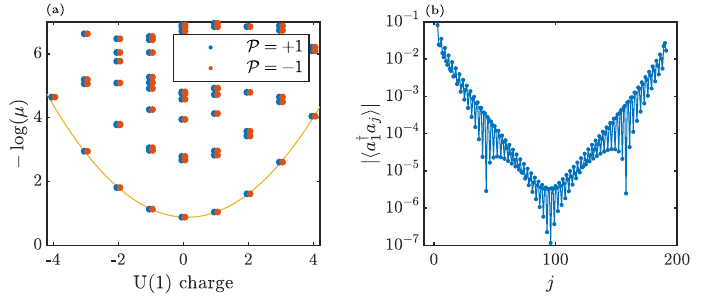}
    \caption{Finite MPS simulations with the conservation of the $\mathcal{P}$ symmetry, for parameters $U_0\!=\!-1.5$, $\alpha\!=\!1/2$, $\nu\!=\!1/3$, and system size $L\!=\!186$. (a) The entanglement spectrum in the middle of the chain, with $\U(1)$ charge and single-leg parity; the solid line is a quadratic fit. (b) The two-point correlation function $\langle\hat{a}_1^\dagger \hat{a}_j\rangle$.}
    \label{fig:fmps_ent}
\end{figure}

 Having established the stability of the topological phase in the thermodynamic limit, we now confirm the topological signatures on a finite system with open boundary conditions. On a finite ladder, a beautiful signature of the Majorana edge modes is provided by the twofold degeneracy in the entanglement spectrum, superimposed by the free-boson boundary CFT spectrum coming from the gapless mode~\cite{Laeuchli2013, Kraus2013}. We plot this characteristic entanglement spectrum in Fig.~\ref{fig:fmps_ent}(a), for a representative point in the topological phase. In addition,  Fig.~\ref{fig:fmps_ent}(b) displays the two-point correlation function $\langle\hat{a}_1^\dagger \hat{a}_j\rangle$, which shows an exponential decay in the bulk and a revival at the end of the chain; this constitutes a direct signature of the MZM in this system.

\section{Concluding remarks}\label{sect:conclusions}

This work introduced a scheme to generate effective pair-hopping processes within a fermionic ladder setting, offering a realistic experimental framework to explore parity-preserving dynamics and topological superconducting phases in quantum-engineered systems. In contrast to previous proposals~\cite{Kraus2013,Tausendpfund_2023}, our Floquet scheme does not rely on any perturbative elimination of auxiliary degrees of freedom (e.g.~internal atomic states), but rather builds on a high-frequency assumption for the driving sequence. 

Our model exhibits a topological ground-state featuring MZMs, which are protected by a reasonable bulk gap ($\Delta_{\rm{topo}}\!\approx\!0.1 \tau$ for bare attractive interactions $U_0\!\approx\!-1.5 \tau$), as well as by the $\mathbb Z_2$ (parity) symmetry inherent to our driving scheme. Majorana edge modes could be finely detected through available spectroscopic schemes~\cite{Kraus_2012, Nascimbene_2013}, and braiding operations could be implemented in an extended network configuration~\cite{Kraus_2013_braiding}. We stress that the Majorana physics (anti-bonding sector) is decoupled from the gapless bosonic mode (bonding sector), such that these two distinct features could be probed independently using well-designed spectroscopic probes, e.g.~Bragg spectroscopy~\cite{Bragg_Ketterle}.

In practice, the ground-state of the effective Hamiltonian could be reached through (quasi) adiabatic quantum state preparation, i.e.~by initializing a simple many-body state on the ladder, and then slowly ramping up the driving sequence; see for instance~\cite{leonard2023realization}. In the present situation, one could start by preparing a trivial insulating state on a single decoupled 1D chain, trapping exactly one fermion on every other site, e.g.~by using a strong staggered potential. Then, one would perform a ($\pi/2$) pulse $\hat P$, which delocalizes the fermions along each rung of the ladder, in view of targeting the anti-bonding sector. Finally, a slow ramping up of the driving sequence would allow one to reach the ground state of the effective Hamiltonian, at filling $\nu\!=\!1/4$. We note that other preparation schemes could be envisaged, such as the recent proposal of Ref.~\cite{michen2024adiabatic}, which exploits an additional magnetic flux.

Beyond ground-state physics, $\mathbb Z_2$-preserving dynamics could also be explored far from equilibrium; we expect these characteristic dynamics to occur on time scales $\sim 1/\epsilon$, where $\epsilon\!=\! \eta - \pi/2$ quantifies the error in the pulse sequence.

Our Floquet scheme requires two main experimental ingredients:~anisotropic nearest-neighbor interactions between spinless fermions on an optical lattice, and dynamical control over nearest-neighbor tunneling matrix elements. Such a setting could be designed by manipulating ultracold dipolar atoms under a quantum gas microscope, as was recently implemented with magnetic erbium atoms~\cite{su2023dipolar}. Importantly, this platform offers fine control over the interactions on the lattice, which can be made long-range, anisotropic and attractive. Alternatively, strong nearest-neighbor attractive interactions could also be engineered using Rydberg dressing, as was realized for fermionic $^6$Li atoms in a 2D optical lattice~\cite{NN_interactions_Bakr}.  Besides, local and dynamical control over tunneling matrix elements was recently demonstrated in an optical superlattice under a quantum gas microscope~\cite{Monika_currents,impertro2024}. It would be interesting to explore generalizations of our scheme to other cold-atom settings, e.g.~a fictitious ladder obtained by combining a double well potential with a synthetic dimension of atomic internal states, hence exhibiting SU(N) interactions along the legs of the ladder~\cite{gorshkov2010two,pagano2014one,Celi_synth_dim}; in this scenario, the driving sequence in Eq.~\eqref{eq: U(T) pulse} would correspond to simply activating the hopping between the two wells in a time-periodic manner~\cite{Goldman_PRXQuantum,Vienna_doublewell_Floquet}.

\textit{Note}: During the revisions, we became aware of the preprint~\cite{lin2025emulating}, which proposes a similar driving scheme to create pair-hopping processes and topological phases in a two-component fermionic system.

\textit{Acknowledgments.---}
We acknowledge Luca Barbiero and Leonardo Mazza for insightful exchanges at various stages of the project. We are also grateful to Alec Douglas, Michal Szurek, and Markus Greiner for their valuable inputs concerning the experimental realization of our scheme. We finally thank Sebastian Diehl, Jean-S\'{e}bastien Caux, Jean Dalibard, Sylvain Nascimbene, Niklas Tausendpfund, Julian L\'{e}onard, Jerome Beugnon, Raphael Lopes for useful discussions and communications.  This work was supported by the ERC (LATIS project), the EOS project CHEQS, the FRS-FNRS Belgium and the Fondation ULB. LPG acknowledges support provided by the L'Or\'eal-UNESCO for Women in Science Programme.

\bibliography{mibib}

% Appendices : 

%\newpage

\vspace{1cm}

\appendix

\section{Pure pair-hopping from a modified driving sequence}\label{app_pair}
We present a modified driving sequence that leads to an effective Hamiltonian with pure pair-hopping, i.e.~an ``ideal" setting that does not include the other two-body terms proportional to $U_2$ in Eq.~\eqref{eq: H_eff ferm. rep.}. In this ``ideal" case, each wire simply acts as a reservoir of Cooper pairs for the other wire, and vice-versa, which is at the essence of a topological phase with MZMs~\cite{Kraus2013}. As detailed below, this modified scheme combines the pulse sequence described in Eq. \eqref{eq: U(T) pulse} with an additional time-modulation of the bare interaction strength $U_0(t)$.

First, we introduce a variant of the decoupled fermionic ladder obtained by simply inverting the sign of the bare interaction strength $U_0$
\begin{align}
    \h H'_0 &= -\tau \sum_j \left( \h a^\dg_j \h a_{j+1} + \h b^\dg_j \h b_{j+1} + \hc \right) \notag \\ 
    &\, - U_0 \sum_j \left( \h a^\dg_j \h a^\dg_{j+1} \h a_{j+1} \h a_j + \h b^\dg_j \h b^\dg_{j+1} \h b_{j+1} \h b_j\right)\, .
    \label{eq: modulated H_0}
\end{align}
Then, in addition to the pulse operated by $\h P \smeq e^{i \eta \h J_x}$, we introduce a second pulse operator chosen in the form $\h W \! \equiv \! e^{i \xi \h J_y}$, which will be used to rotate $\h H'_0$ within each driving period. Here, we fix the drive parameters $\eta\!=\!\xi\!=\!\pi/2$ to ensure the $\mathbb Z_2$-symmetry. We now design a specific driving sequence generated by the following time-evolution operator over one driving period
\begin{align}
    \h U(T) &= \h W^\dg e^{-i \alpha_4 T \h H'_0} \h W e^{-i \alpha_3 T \h H_0} \h P^\dg e^{-i \alpha_2 T \h H_0} \h P e^{-i \alpha_1 T \h H_0} \notag \\ 
    &= e^{-i \alpha_4 T \h H_2} e^{-i \alpha_3 T \h H_0} e^{-i \alpha_2 T \h H_1} e^{-i \alpha_1 T \h H_0} \, , \label{eq: U(T) two pulses}
\end{align}
where $\h H_2 \! \equiv \! \h W^\dg \h H'_0 \h W$, and where $\alpha_1$, $\alpha_2$, $\alpha_3$ and $\alpha_4$ are drive parameters such that $\sum_i \alpha_i \smeq 1$, $\alpha_i \! > \! 0$. 

Applying the Trotter approximation to Eq.~\eqref{eq: U(T) two pulses}, we obtain the effective (Floquet) Hamiltonian 
\begin{equation}
\h H_{\rm{eff}} \smeq (\alpha_1 + \alpha_3) \h H_0 + \alpha_2 \h H_1 + \alpha_4 \h H_2 + O(T).\label{eq_app_H_eff}
\end{equation}
In view of clarifying how the processes generated by $\h H_2$ can cancel the ``undesired" terms generated by $\h H_1$, we write $\h H_1$ and $\h H_2$ in the spin representation 
\begin{align}
    \begin{cases}
        \h H_1 = e^{-i \frac{\pi}{2} \h J_x} \h H_0 e^{i \frac{\pi}{2} \h J_x} = \frac{U_0}{2} \sum_j \left( \h N^j \h N^{j+1} + 4 \h J_y^j \h J_y^{j+1} \right) \\ 
        \h H_2 = e^{-i \frac{\pi}{2} \h J_y} \h H'_0 e^{i \frac{\pi}{2} \h J_y} = -\frac{U_0}{2} \sum_j \left( \h N^j \h N^{j+1} + 4 \h J_x^j \h J_x^{j+1} \right)\, .
    \end{cases} \label{eq: H1 H2 spin rep}
\end{align}
For the sake of clarity, we omitted the intra-wire single-particle hopping in Eq.~\eqref{eq: H1 H2 spin rep}, which contributes equally to both Hamiltonians $\h H_1$ and $\h H_2$. The expression of $\h H_2$ in Eq.~\eqref{eq: H1 H2 spin rep} can be derived using the Baker-Campbell-Hausdorff formula, in the same manner as we obtained $\h H_1$ through Eq.~\eqref{eq: BCH} in the main text. 

From the definition of the spin operators in Eq.~\eqref{eq: spin op.}, one readily finds that a proper combination of $\h J_y^j \h J_y^{j+1}$ and $\h J_x^j \h J_x^{j+1}$ yields the desired pair-tunneling process:
\begin{equation}
\h J_y^j \h J_y^{j+1} - \h J_x^j \h J_x^{j+1} = - \frac{1}{2} \left (\h a^\dg_j \h a^\dg_{j+1} \h b_{j+1} \h b_j + \hc \right ).\label{eq:pair_hop_spin}
\end{equation}

Hence, combining Eqs.~\eqref{eq_app_H_eff}-\eqref{eq:pair_hop_spin} and setting $\alpha_2\!=\!\alpha_4$, one finally obtains the target effective Hamiltonian
\begin{align}
    &\h H_{\rm{eff}} = -\tau \sum_j \left( \h a^\dg_j \h a_{j+1} + \h b^\dg_j \h b_{j+1} + \hc \right) \label{eq: Heff U_0(t)}\\ 
    &\, + U_n \sum_j \left( \h n^a_j \h n^a_{j+1} + \h n^b_j \h n^b_{j+1} \right) + U_p \sum_j \left( \h a^\dg_j \h a^\dg_{j+1} \h b_{j+1} \h b_j + \hc \right) ,\notag 
\end{align}
where $U_n \smeq U_0 (\alpha_1 + \alpha_3)$ and $U_p \smeq - U_0 \alpha_2$. 

 As a final remark, we note that our scheme assumes identical wires, with same tunneling matrix elements $\tau\!=\!\tau_a\!=\!\tau_b$. In case irregularities are present, $\tau_a\!\ne\!\tau_b$, one could include additional ``spin-echo" $\pi$ pulses in the sequence to effectively annihilate the resulting (undesired) effects.

\section{Driving protocol with a continuous drive}\label{app_cont}
We present an alternative protocol to the pulse sequence scheme described in the main text, replacing the pulses by a continuous drive. We demonstrate below that the same effective (Floquet) Hamiltonian [Eq.~\eqref{eq: H_eff ferm. rep.}] can be obtained with this approach, up to renormalized coefficients and without any fined-tuned drive parameter. 

The time-dependent Hamiltonian of the driven ladder is written in the form
\begin{align}
    \h H(t) &= \h H_0 + \h V(t) \notag \\ 
    &= \h H_0 + A \cos (\omega t) \h J_x\, ,
\end{align}
where we recall that $\h H_0$ describes the decoupled-wire setting in Eq.~\eqref{eq: H_0}, and where $\h J_x$ is a spin operator [Eq.~\eqref{eq: spin op.}] describing inter-wire single-particle hopping. From an experimental perspective, the primary challenge of this approach concerns the need to switch the sign of the (time-modulated) tunneling matrix elements during the protocol~\citep{Pieplow2018}.

In the high-frequency limit, one can derive an effective Hamiltonian as an expansion in powers of the driving period $T \!=\! 2\pi/\omega$ \cite{Goldman_Dalibard_PRX}:
\begin{align}
    \h H_{\rm{eff}} = \h H_0 + \frac{1}{\omega} \sum_{j=1}^{\infty} \frac{1}{j} \left[ \h V^j, \h V^{-j} \right] + O(1/\omega^2)\, , \label{eq: Eff Ham expansion}
\end{align}
where the operators $\h V^j$ are the Fourier components of the time-periodic potential, i.e.~$\h V(t) \!=\! \sum_{j=1}^{+\infty} \left\{ \h V^j e^{ij \omega t} + \h V^{-j} e^{-ij \omega t} \right\}$. In order to induce novel 2-body processes at the lowest order, we operate within a strong driving regime~\cite{Goldman_PRA_resonant}, characterized by $A \! = \! \omega K_0 \! \gg \! \tau,U_0$,  where $K_0\!\sim\!1$. In this framework, it is essential to choose a proper reference frame to ensure the convergence of the expansion~\eqref{eq: Eff Ham expansion}.
 In this case, the proper change of basis is provided by the unitary transformation
\begin{align}
&\h R(t) \! =\! e^{ i K_0 \sin(\omega t) \h J_x},\label{eq:changeframe}
\end{align}
and the time-dependent Hamiltonian is transformed according to
\begin{align}
& \hat H (t) \longrightarrow \hat{\mathcal H} (t)=\h R(t)\hat H (t)\h R^{\dagger}(t) - i \h R(t) \partial_t \h R^{\dagger}(t) \notag\\
&\hspace{2.32cm}=\h R(t)\hat H_0\h R^{\dagger}(t) \notag \\
&\hspace{2.32cm} \equiv \hat{\mathcal H}_0 + \hat{\mathcal V} (t), \label{eq:changeframe_2}
\end{align}
where we introduced the static ($\hat{\mathcal H}_0$) and time-dependent ($\hat{\mathcal V}$) components of the Hamiltonian in the moving frame. Following Refs.~\cite{Goldman_Dalibard_PRX,Goldman_PRA_resonant}, we compute the effective Hamiltonian in this moving frame,
\begin{equation}
\hat{\mathcal H}_{\rm{eff}} = \hat{\mathcal H}_0 + \frac{1}{\omega} \sum_{j>0} \frac{1}{j} [\hat{\mathcal V}_j , \hat{\mathcal V}_{-j}] + \mathcal{O} (1/\omega^2),\label{Heff_Ham_general}
\end{equation}
where we introduced the Fourier components in the moving frame, $\hat{\mathcal V} (t)\!=\!\sum_{j\ne0}\hat{\mathcal V}_j e^{i j\omega t}$. We point out that the $1/\omega$-expansion in Eq.~\eqref{Heff_Ham_general} converges in the high-frequency limit, even in the strong-driving regime $K_0\!\sim\!1$, thanks to the well-chosen moving frame defined in Eqs.~\eqref{eq:changeframe}-\eqref{eq:changeframe_2}.

According to Eq.~\eqref{Heff_Ham_general}, the lowest-order effective Hamiltonian is simply given by
\begin{align}
  \hat{\mathcal H}_{\rm{eff}} = \frac{1}{T} \int_0^T \h R(t)\hat H_0\h R^{\dagger}(t) \, dt \, + \mathcal{O} (1/\omega).\label{H_eff_lowest_order}
\end{align}
Considering the fermionic representation, the effective Hamiltonian in Eq.~\eqref{H_eff_lowest_order} explicitly reads
\begin{align}
    \hat{\mathcal H}_{\rm{eff}} &=-\tau \sum_j \left( \h a^{\dg}_j \h a_{j+1} + b^{\dg}_j \h b_{j+1} + \hc \right) \label{eq: H_eff cont. drive} \\ 
    &\, + \tilde U_1 \sum_j \left( \h n^a_j \h n^a_{j+1} + \h n^b_j \h n^b_{j+1} \right) + \tilde U_2 \sum_j \left( \h n^a_j \h n^b_{j+1} + \h n^b_j \h n^a_{j+1} \right) \notag \\ 
    &\, + \tilde U_2 \sum_j \left( \h a^{\dg}_j \h b^{\dg}_{j+1} \h a_{j+1} \h b_j - \h a^{\dg}_j \h a^{\dg}_{j+1} \h b_{j+1} \h b_j + \hc \right) \notag
\end{align}
where $\tilde U_1 \! = \! \frac{U}{4} \left( 3+ \mathcal{J}_0(2K_0) \right)$ and $\tilde U_2 \! = \! \frac{U}{4} \left( 1-\mathcal{J}_0(2K_0) \right)$, and $\mathcal{J}_0$ is a Bessel function of the first kind. As anticipated, this protocol generates the same 2-body processes as in Eq.~\eqref{eq: H_eff ferm. rep.}, with renormalized couplings. In particular, the effective Hamiltonian derived in Eq.~\eqref{eq: H_eff cont. drive} exhibits the desired parity ($\mathbb Z_2$) symmetry, for any values of the driving parameters. 

Interestingly, the first-order corrections to the effective Hamiltonian [Eq.~\eqref{Heff_Ham_general}] are identically zero in this scheme. This result can be obtained by identifying the Fourier components of the time-dependent operator $\h{\mathcal V}(t)$ in the moving frame as 
\begin{align}
    &\h{\mathcal V}_{\pm j} = \alpha_j \h O_1 \pm \beta_j \h O_2 , \quad j > 0, \\ 
    &\ =\sum_k \left[ \alpha_j\left( \h J_z^k \h J_z^{k+1} - \h J_y^k \h J_y^{k+1} \right) + \beta_j \left( \h J_y^k \h J_z^{k+1} + \h J_z^k \h J_y^{k+1} \right) \right] \, , \notag 
\end{align}
where we chose the spin representation for the sake of clarity. The coefficients $\alpha_j$ and $\beta_j$ are explicitly given by
\begin{align}
    \begin{cases}
        \alpha_j \smeq \frac{1}{2} \left( \mathcal{J}_j (2K_0) \smpl \mathcal{J}_{-j}(2K_0)\right) \\ 
        \beta_j \smeq \frac{1}{2i} \left( \mathcal{J}_j (2K_0) \smmi \mathcal{J}_{-j}(2K_0)\right) \, .
    \end{cases} 
\end{align} 
According to Eq.~\eqref{Heff_Ham_general}, the first-order corrections to the effective Hamiltonian are determined by the commutators of the Fourier components, $\left[ \h{\mathcal V}_j, \h{\mathcal V}_{-j} \right]$, which are found to vanish for each $j\!>\!0$,
\begin{align}
    \left[ \h{\mathcal V}_j, \h{\mathcal V}_{-j} \right] = 2 \alpha_j \beta_j \left[ \h O_1, \h O_2 \right] = 0 \, ,
\end{align}
where we used the property $\mathcal J_{-j}(x)\!=\! (-1)^j \mathcal J_j(x)$ satisfied by the Bessel function of the first kind. \\ 

The second-order corrections can be evaluated by making use of the formulas derived in Ref.~\cite{Goldman_Dalibard_PRX}, yielding additional 2-body processes (including $\Z_2$-breaking terms) of order $\tau U_0^2/\omega^2$ and $U_0^3/\omega^2$, which can thus be safely neglected in the high-frequency regime. Under realistic experimental conditions, one would consider bare interactions and hopping amplitudes of order $U_0, \tau\!\sim\!$ 100 Hz ($\hbar\!=\!1$), and a driving frequency of order $\omega\!\sim\!$ 1000 Hz; in this case, the topological gap would be of order $\Delta_{{\rm topo}}\!\sim\!$ 10 Hz, namely, an order of magnitude larger than these second-order corrections. \\

Finally, we remark that the moving-frame operator $\h R(t)$ introduced in Eq.~\eqref{eq:changeframe} describes the micromotion in the original (``lab") frame. Indeed, the time-evolution operator in the original frame takes the form~\cite{Goldman_PRA_resonant}
\begin{equation}
\hat U (t; t_0)=\h R^{\dagger}(t) e^{-i (t-t_0) \hat{\mathcal H}_{\rm{eff}}} \h R(t_0),
\end{equation}
where we have neglected the small micromotion within the moving frame, i.e.~the kick operator $\hat{\mathcal K}(t)\!\approx\!\hat 1$; see Ref.~\cite{Goldman_PRA_resonant}. One deduces from Eq.~\eqref{eq:changeframe} that the micromotion observed in the lab frame is generated by single-particle inter-chain hopping processes, which break the parity ($\mathbb Z_2$) symmetry within each period of the drive. Since $\h R(0)\!=\!\h R(T)\!=\!\hat 1$, one finds that the time-evolution operator describing  stroboscopic dynamics \emph{in the lab frame} simply reads
\begin{equation}
\hat U (t_N; 0)=e^{-i t_N \hat{\mathcal H}_{\rm{eff}}}, \quad t_N\!=\! N T, \quad N \text{ integer},
\end{equation}
where $\hat{\mathcal H}_{\rm{eff}}$ is the effective Hamiltonian in Eq.~\eqref{eq: H_eff cont. drive}, derived in the moving frame.\\

\section{Infinite MPS simulations}\label{app_MPS}

In this Appendix, we provide more details on the numerical simulations performed using the formalism of infinite matrix product states (MPS). The approach is different than previous works~\cite{Kraus2013, Iemini2017, Lisandrini2022, Tausendpfund_2023}, where finite MPS were used for detecting signatures of MZMs in number-conserving ladder systems.

\textit{Kitaev chain.---}%
First, let us start with the simple Kitaev chain \cite{Kitaev2001}, a chain of spinless fermions with the Hamiltonian
\begin{multline}
    \hat{H}_{\text{chain}} = -t \sum_j \left(\h{c}_j^\dagger \h{c}_{j+1} + \h{c}_{j+1}^\dagger \h{c}_{j}  \right) - \mu \sum_j \h{c}_j^\dagger \h{c}_j \\ - \Delta \sum_j \left( \h{c}_j \h{c}_{j+1} + \h{c}_{j+1}^\dagger \h{c}_{j}^\dagger \right) .
\end{multline}
The model exhibits a topologically trivial phase for $|\mu|>2t$ and a topologically non-trivial phase for $|\mu|<2t$, with a phase transition at the points $\mu\!=\!\pm2t$. We can simulate this phase diagram directly in the thermodynamic limit by representing the ground state as an infinite MPS \cite{Vanderstraeten2019}. Two options are available for encoding the statistics of the fermions: either we perform a Jordan-Wigner transformation to map the Kitaev chain to the transverse-field Ising model, or we use the formalism of fermionic MPS \cite{Bultinck2017, Mortier2024} to represent MPS directly in the fermionic basis. The latter option requires us to explicitly encode the fermionic parity symmetry $\mathcal{P}$
\begin{equation}
    \hat{\mathcal{P}} = (-1)^{\sum_j \hat{c}_j^\dagger \hat{c}_j },
\end{equation}
into the MPS tensor, so that all numerical MPS algorithms can keep track of the anticommutation relations of the fermionic degrees of freedom. The first option does not imply this requirement, so the infinite MPS is allowed to break the $\Z_2$ symmetry of the Ising model.
\par When we optimize an infinite MPS approximation in the topological phase, the features of the resulting state depends on whether we use bosonic or fermionic MPS. Indeed, it is well-known that the topological phase of the Kitaev chain maps to the symmetry-broken phase of the Ising model \cite{Greiter2014}; since infinite MPS always favour the states with maximal symmetry breaking, our infinite bosonic MPS will yield one of the symmetry-broken states with a non-zero local order parameter. With fermionic MPS, however, we find an MPS with tensors that can be decomposed into the form
\begin{equation}
    A^0 = \begin{pmatrix} B^0 & 0 \\ 0 & B^0 \end{pmatrix}, \quad A^1 = \begin{pmatrix} 0 & B^1 \\ - B^1 &  0 \end{pmatrix} \;,
\end{equation}
with two smaller matrices $B^0$ and $B^1$. For a bosonic MPS, this form would imply that the MPS can be further decomposed as the global superposition of two different MPS. For a fermionic MPS, however, such a decomposition is not possible since this would break the fermionic parity symmetry of the tensors. For that reason, this MPS is said to be ``irreducible'' \cite{Bultinck2017}, although it is in fact a global superposition of two states that break the fermionic parity symmetry. This form of the fermionic MPS is characteristic of the topological order in the Kitaev chain, because it leads directly to (i) exact twofold degeneracies in the entanglement spectrum \cite{Fidkowski2010, Turner2011}, (ii) a twofold degeneracy when the MPS tensors are put on a finite chain with open boundary conditions, and (iii) a unique MPS with odd parity on a chain with periodic boundary conditions \cite{Bultinck2017}.
\par With infinite MPS, therefore, using either fermionic MPS for the original model or bosonic MPS for the mapped Ising model gives quite a different picture. In some sense, it is a lot more natural to allow the system to break the $\Z_2$ symmetry explicitly in the bosonic version, instead of working with these special MPS forms in the fermionic version. The bond dimension of the fermionic MPS is artificially doubled by insisting on the fermionic parity symmetry, so the bosonic version is also a lot more efficient from the computational point of view. Finally, in the bosonic Ising model the symmetry breaking is characterized by a local order parameter that can be simply evaluated for the infinite MPS, whereas any order parameter in the fermionic Kitaev chain is necessarily described by a non-local operator.
\par We should note that for finite systems, the situation between the bosonic and the fermionic approach is quite similar. For a finite system, the $\Z_2$ symmetry remains unbroken in the bosonic case as well, and the entanglement spectrum on a finite system shows the same degeneracies as in the fermionic case \cite{Greiter2014} -- in that sense, the degeneracies in the entanglement spectrum are not a signature of topological order.

\textit{Ladder.---}%
Given these insights, let us now motivate our approach for the ladder system. We first take the simplest version as it appeared in Ref.~\cite{Kraus2013}:
\begin{multline} \label{eq:kraus}
    \hat{H}_{\text{ladder}} = -t \sum_j \left(\hata_j^\dagger \hata_{j+1} + \hata_{j+1}^\dagger \hata_{j} + \hatb_j^\dagger \hatb_{j+1} + \hatb_{j+1}^\dagger \hatb_{j} \right) \\ + W \sum_j \left( \hata_j^\dg \hata_{j+1}^\dg \hatb_j \hatb_{j+1} + \hatb_j^\dg \hatb_{j+1}^\dg \hata_j \hata_{j+1} \right) \; ,
\end{multline}
taking the notation for fermionic creation and annihilation operators from the main text. The first symmetry of the model is the total fermionic charge, generated by the total number operator
\begin{equation}
    \h{N}_t = \h{N}_a + \h{N}_b, \qquad \h{N}_a = \sum_j \hata_j^\dg \hata_j,  \quad \h{N}_b= \sum_j  \hatb_j^\dg \hatb_j ,
\end{equation}
which also allows us to define the total fermionic parity $\hat{\mathcal{P}}_t = (-1)^{\hat{N}_t}$. In addition, there is the single-leg parity operator
\begin{equation}
    \hat{\mathcal{P}} = (-1)^{\hat{N}_a} = \pm (-1)^{\hat{N}_b} .
\end{equation}
The physics of the topological phase of this ladder system is now related to the spontaneous breaking of this single-leg parity symmetry, whereas the total fermionic charge $N_t$ and parity $\mathcal{P}_t$ remain good quantum numbers. So, in contrast to the Kitaev chain, the single-leg parity does not take the role of a superselection rule. We can, therefore, use infinite fermionic MPS with the explicit conservation of $\hat{N}_t$ and $\hat{\mathcal{P}}_t$, but without imposing the single-leg parity $\hat{\mathcal{P}}$. Upon optimizing an infinite MPS $\ket{\Psi_0}$ for this model in the topological phase, we can therefore expect that an optimized MPS breaks the single-leg parity spontaneously. We can, however, not characterize this symmetry breaking by a local order parameter, but instead directly monitor to what extent the MPS is not an eigenstate of the parity operator:
\begin{equation}
    \lambda = - \lim_{N\to\infty} \frac{1}{N} \log \left| \bra{\Psi_0} \hat{\mathcal{P}} \ket{\Psi_0} \right|,
\end{equation}
where have formally introduced the diverging system size $N$ to define a ``log-fidelity density''.
\par The approach of symmetry-breaking infinite MPS is particularly instructive, because it also allows us to estimate the topological gap efficiently. Note that the topological gap is not directly accessible in the energy spectrum of the Majorana ladder on a system with open boundary conditions, because of the presence of the gapless sector. Instead, one should consider the system on periodic boundary conditions, where the topological gap is given by 
\begin{equation}
    \Delta_{{\rm topo}}  = \frac{1}{2}\Big(\Delta_{Q=+1}+\Delta_{Q=-1}\Big),
\end{equation}
with $\Delta_Q$ the excitation energy of the lowest lying excited state with a fermion charge $Q$ relative to the ground state charge \cite{Kraus2013}. In the infinite system, the gapless mode immediately yields a continuum of states in the spectrum, but we can also access the topological bulk gap in the charged sector. %See Fig.~\ref{fig:spectrum} for a schematic overview of the spectrum in these three different situations.

%\begin{figure}[t]
%    \centering
%    \includegraphics[width=0.9999\columnwidth]{Figures/spectrum.pdf}
%    \caption{The qualitative energy spectrum of the Kitaev ladder for different boundary conditions.}
%    \label{fig:spectrum}
%\end{figure}

\par
\begin{figure}
    \centering
    \includegraphics[width=0.9999\columnwidth]{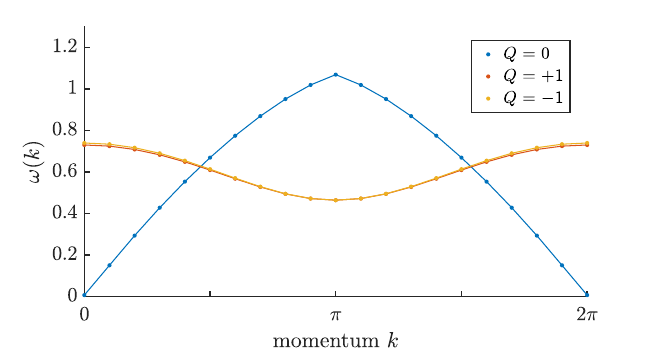}
    \caption{The dispersion $\omega(k)$ as a function of momentum $k$ of the lowest-lying excitation energy in three different charge sectors, for the model in Eq.~\eqref{eq:kraus} with $W=-1.8$ and filling $\nu=1/3$. We have shifted the chemical potential towards the middle of the gap between the $Q=\pm1$ sectors.}
    \label{fig:disp}
\end{figure}

 The excitation energies $\Delta_{Q=\pm1}$ are accessible by using the MPS excitation ansatz, a variational parametrization of excited states on top of an infinite MPS with well-defined momentum and charge \cite{Haegeman2012, ZaunerStauber2018b, Vanderstraeten2019}. Applying this ansatz, we can find the lowest-lying excitations in the $Q\!=\!0$ and the $Q\!=\!\pm1$ sectors as a function of momentum, and find the minimum of these dispersion relations as estimates for $\Delta_{Q=0}$ and $\Delta_{Q=\pm1}$. As an illustration, our numerical results for the dispersions are shown in Fig.~\ref{fig:disp}.
\par
\begin{figure}[t]
    \centering
    \includegraphics[width=0.9999\columnwidth]{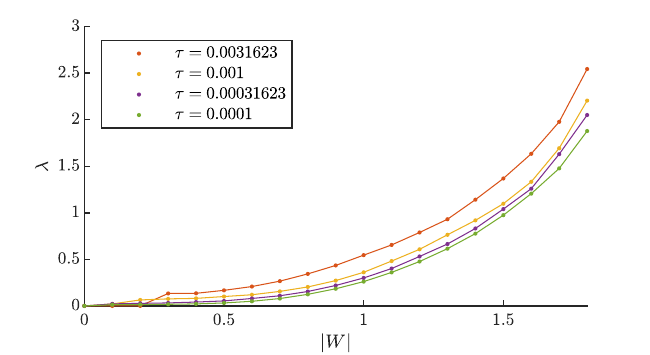}
    \includegraphics[width=0.9999\columnwidth]{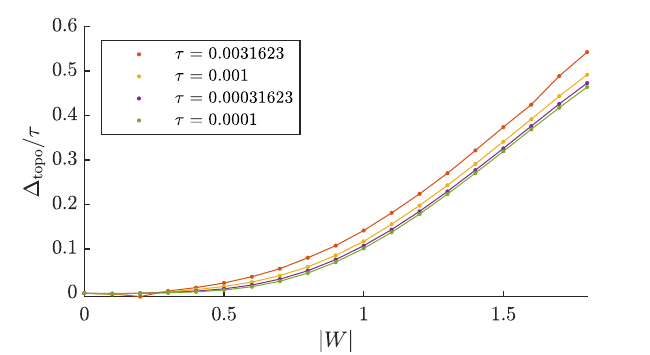}
    \caption{Symmetry breaking (top) and topological gap (bottom) as a function of $W$ for a filling of $\nu=1/3$, for different values of the truncation threshold $\tau$.}
    \label{fig:gap_kraus}
\end{figure}

 Using this approach, we can now monitor the symmetry breaking and the topological gap as a function of $W$, the strength of the pair hopping term in Eq.~\eqref{eq:kraus}. The numerical results are given in Fig.~\ref{fig:gap_kraus}. We show the results for different truncation thresholds $\tau$ in the MPS -- i.e., different bond dimensions. From these plots, we can see that both the symmetry breaking parameter $\lambda$ and the topological gap converge nicely as a function of truncation threshold. We observe that the gap opens up exponentially slowly, which is in agreement with a bosonization analysis \cite{Cheng2011}: starting from two decoupled non-interacting wires with effective Luttinger parameter $K_{a,b}=1$, the pair hopping term is marginal and opens up the gap according to an exponential.

\begin{figure}
    \centering
    \includegraphics[width=0.9999\columnwidth]{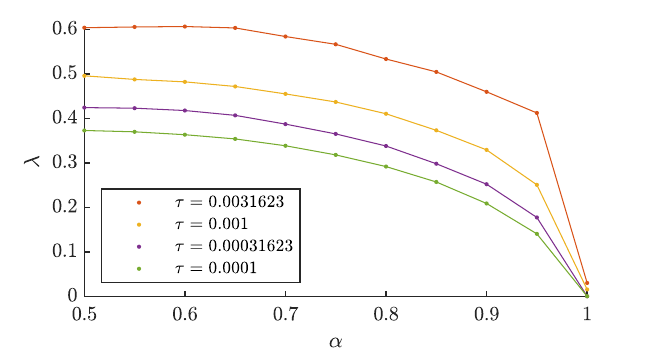}
    \includegraphics[width=0.9999\columnwidth]{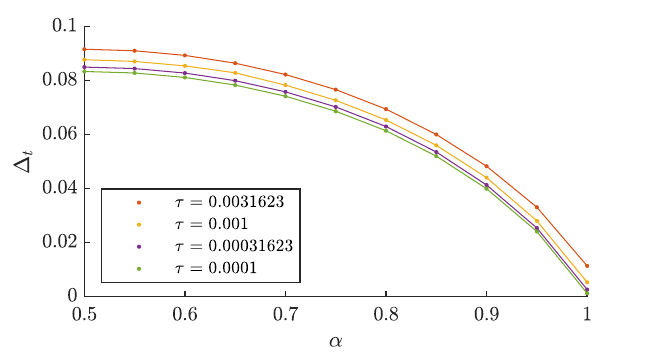}
    \caption{Symmetry breaking (top) and topological gap (bottom) as a function of $\alpha$ for a bare  interaction $U_0\!=\!-1.5$ and filling of $\nu\!=\!1/3$, for different values of the truncation threshold $\tau$.}
    \label{fig:gap_alpha_ext}
\end{figure}

\begin{figure}
    \centering
    \includegraphics[width=0.9999\columnwidth]{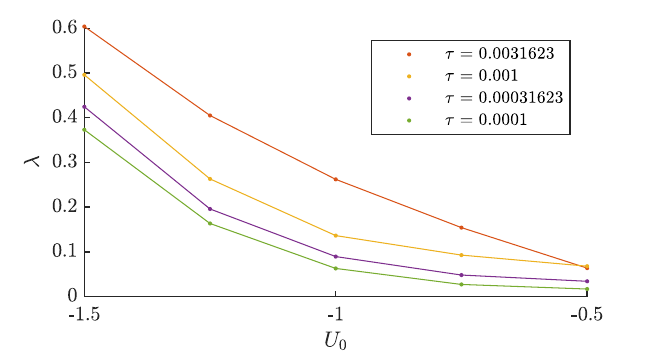}
    \includegraphics[width=0.9999\columnwidth]{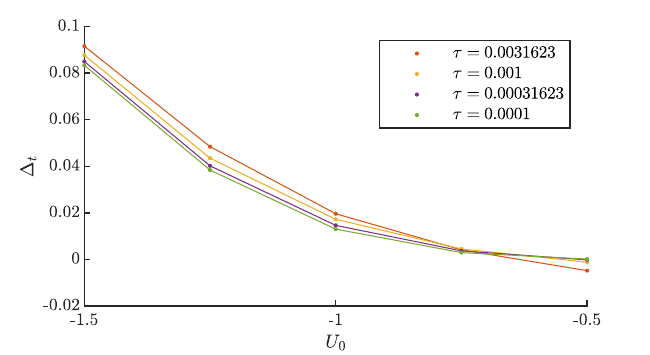}
    \caption{Symmetry breaking (top) and topological gap (bottom) as a function of $U_0$ for a fixed value of $\alpha=0.5$ and a filling of $\nu=1/3$, for different values of the truncation threshold $\tau$.}
    \label{fig:gap_U0_ext}
\end{figure}

\textit{Effective model.---}%
Finally, we use our approach for the effective Hamiltonian in Eq.~\eqref{eq: H_eff ferm. rep.} with parameters $U_0$ (the interchain interactions) and $\alpha$ (the intrachain processes). In Figs.~\ref{fig:gap_alpha_ext} and \ref{fig:gap_U0_ext} the results are shown for the symmetry breaking and the topological gap. The results for the largest bond dimension in this figure were also reported in the main text.

\section{Effects of finite pulse duration and impure pulses}\label{app:pulse}

The pulses entering the sequence in Eq.~\eqref{eq: U(T) pulse} were assumed to be instantaneous. However, in a practical implementation, the single-particle inter-chain hopping processes would be activated during a finite duration $t_p\!\ll\!T$, and with a finite strength, which are generally limited by experimental constraints. In this Appendix, we evaluate the impact of a finite pulse duration $t_p$ on the effective Hamiltonian describing the long-time dynamics of our setting. Our analysis is performed in two steps: (i) we first introduce a finite pulse duration in the sequence, assuming that only inter-chain hopping processes are activated during each pulse; (ii) we then relax this ``pure pulse" hypothesis, by examining the effects of having the bare Hamiltonian $\hat{H}_0$ active during each pulse.

\subsection{Finite pulse duration}

Considering a finite pulse duration $t_p$, the driving sequence now takes the following form
\begin{align}
    \h U(T) &= \h P_{\mathrm{II}} \, e^{-i ((1-\alpha)T - t_p) \h H_0} \, \h P_{\mathrm{I}} \, e^{-i (\alpha T - t_p)\h H_0} \notag \\ 
    &= e^{-i ((1-\alpha)T - t_p) \h H_1} \, e^{-i (\alpha T - t_p) \h H_0} \label{eq: realistic U(T)}\\ 
    &= e^{-iT \heff} \notag
\end{align}
where the pulse operators are explicitly given by
\begin{align}
&\begin{cases}
    \h P_{\mathrm{I}} \equiv  e^{-i t_p J_{\mathrm I} \h J_x} = e^{i \frac{\pi}{2} \h J_x} \\ 
    \h P_{\mathrm{II}} \equiv e^{-i t_p J_{\mathrm{II}} \h J_x} = e^{-i \frac{\pi}{2} \h J_x} = e^{i \frac{3\pi}{2} \h J_x} \, .
\end{cases}
\end{align}
Here, we explicitly set $\eta\!=\!\pi/2$ and we introduced the inter-chain hopping strength
\begin{equation}
 J_{\mathrm{II}} \smeq -\frac{3\pi}{2 t_p} \smeq 3J_{\mathrm I}. 
 \end{equation}
 As in the main text, $\h H_1 = e^{-i \frac{\pi}{2} \h J_x} \h H_0\, e^{i \frac{\pi}{2} \h J_x}$, and we thus recover the same time-evolution operator as in Eq.~\eqref{eq: U(T) pulse} in the limit $t_p \to 0$. 
 
In this framework, the effective Hamiltonian now takes the form 
\begin{equation}
\heff \smeq (\alpha - \frac{t_p}{T}) \h H_0 + \left[ 1-\alpha - \frac{t_p}{T} \right] \h H_1,\label{H_eff_finite}
\end{equation}
at lowest order in the period $T$. 

In the main text, we found that the topological gap was maximized when the effective Hamiltonian was of the form
\begin{equation}
\heff^{{\rm opt}} \smeq \tilde \alpha \left( \h H_0 + \h H_1 \right).
\end{equation}
In the case $t_p\!=\!0$, this amounts to setting $\tilde \alpha\!=\!\alpha\!=\!1/2$. In the present case of finite pulse duration, this optimal regime is reached by setting $\tilde \alpha  \smeq (1/2 - t_p/T)$. 

Comparing $\heff^{{\rm opt}}$ with the effective Hamiltonian derived in the main text (with the optimal values $\eta\!=\!\pi/2$ and $\alpha\!=\!1/2$), we thus deduce that a finite pulse duration $t_p$ leads to a rescaling of energies by a factor $1 - 2t_p/T$.

\begin{figure}[htbp!]
    \centering
\includegraphics[width=0.999\columnwidth]{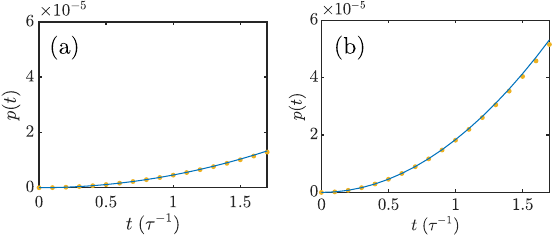}
    \caption{(a)-(b) Time evolution of the mean probability of changing parity ($Z_2$ breaking), considering two fermions on a single plaquette (four lattice sites). In each panel, we compare the exact dynamics generated by the full time-dependent Hamiltonian \eqref{eq: exact Ham impure pulse} (blue curve), with the effective Hamiltonian \eqref{eq:Heffimpurepulse} (yellow dots). Both dynamics are evaluated at stroboscopic times ($t_n \!=\! n, T$, $n \!\in\! \Z$). The pulse duration is set to: (a) $t_p/T \!=\! 1/40$, and (b) $t_p/T \!=\! 1/20$. The driving parameters are $T=0.1$, $\alpha \!=\! 1/2$, and $\eta \!=\! \pi/2$, and the bare system parameters are $U_0 \!=\! -0.7$ and $\tau \!=\! 1$.}
    \label{fig: realistic pulse}
\end{figure}

\subsection{Effects of impure pulses}

The time-evolution operator in Eq.~\eqref{eq: realistic U(T)} amounts to neglecting the action of the bare Hamiltonian $\h H_0$ during the pulses. This is justified since the activated inter-chain hopping strength dominates over all other energy scales within each pulse. We now relax this ``pure pulse" hypothesis and determine the first-order corrections to the effective Hamiltonian.\\ 

In this framework, we consider that the system is described by a time-dependent Hamiltonian of the form
\begin{align}
    \h H(t) = \h H_0 + A \, f(t) \,\h J_x , \label{eq: exact Ham impure pulse}
    \end{align}
where the pulse function describes a square  drive [see Fig.~\ref{fig_sketch}] 
\begin{align} f(t)= \begin{cases}
        1 \quad \mathrm{if} \ \alpha T - t_p \le t \le \alpha T\, ; \\ 
        3 \quad \mathrm{if} \ T-t_p \le t \le T \, ;\\
        0 \quad \mathrm{otherwise} \, ,
    \end{cases} \notag 
\end{align}
and where the coefficient $A$ satisfies the condition $A \, t_p\!=\!\eta\!=\!\pi/2$.

Following the same procedure as in Appendix~\ref{app_cont}, we derive an effective Hamiltonian in the high-frequency limit upon performing a proper change of basis. The corresponding unitary transformation $\h R(t)$ is chosen such that the time-dependent Hamiltonian becomes $\h{\mathcal H}(t)\!=\! \h R(t) \h H_0 \h R^\dg(t)\!\equiv\! \h{\mathcal H}_0 + \h{\mathcal V}(t)$ in the moving frame; see Refs.~\cite{Goldman_Dalibard_PRX,Goldman_PRA_resonant}. Here, this is achieved by setting $\h R(t) \smeq e^{i A  \h J_x \int^t f(t') dt'}$. One can then extract the lowest-order effective Hamiltonian by averaging out the time-dependent components [Eqs.~\eqref{Heff_Ham_general}-\eqref{H_eff_lowest_order}]
\begin{align}
  \hat{\mathcal H}_{\rm{eff}} =  \h{\mathcal H}_0 &= \frac{1}{T} \int_0^T \h{\mathcal H}(t) \, dt \, .
\end{align}

The calculation of this effective Hamiltonian can be performed explicitly by splitting the integral as
\begin{align}
    \hat{\mathcal H}_{\rm{eff}} &=  \frac{1}{T}\int_0^{\alpha T-t_p} \h{\mathcal H}(t) dt + \frac{1}{T}\int_{\alpha T - t_p}^{\alpha T} \h{\mathcal H}(t) dt \\   
    &+\frac{1}{T}\int_{\alpha T}^{T-t_p} \h{\mathcal H}(t) dt + \frac{1}{T}\int_{T-t_p}^T \h{\mathcal H}(t) dt \, .
\end{align}
Over the time intervals $[0, \alpha T - t_p]$ and $[\alpha T, T-t_p]$, the time-dependent unitary transformation reduces to $\h R(t)\!=\! \h{\mathbb I}$ and $\h R(t)\!=\! e^{i \eta \h J_x}$, respectively. 
Therefore, the contribution of those intervals to the effective Hamiltonian is the same as in Eq.~\eqref{H_eff_finite}. 
In the two remaining intervals, $\alpha T-t_p \leq t \leq \alpha T$ and $T -t_p \leq t \leq T$, the unitary transformation yields $\h R(t) = e^{iA(t-\alpha T+t_p)\h J_x}$ and $\h R(t) = e^{i(\eta +3A(t-T+t_p))\h J_x}$, respectively. Making use of Eq.~\eqref{eq: BCH} upon evaluating the remaining integrals, and combining all the contributions, we finally obtain the following expression for the effective Hamiltonian
\begin{align}
    \hat{\mathcal H}_{\rm{eff}} &= \left( \alpha - \frac{t_p}{T} \right) \h H_0 + \left( 1 - \alpha - \frac{t_p}{T} \right) \h H_1 \notag \\ 
    &\ + 2 \frac{t_p}{T} \left[ \h H_0 + U_0 \sum_j\left( \h J_y^j \h J_y^{j+1} - \h J_z^j \h J_z^{j+1} \right) \right] \notag \\ 
    &\ + \frac{4\, U_0}{3\pi} \frac{t_p}{T} \sum_j \left( \h J_y^j \h J_z^{j+1} + \h J_z^j \h J_y^{j+1} \right) \, . \label{eq:Heffimpurepulse}
\end{align}
Importantly, one finds that the finite duration of the pulse can generate $\Z_2$-breaking interactions of order $O(U_0 \, t_p/T)$, at the lowest order of the $(1/\omega)$-expansion. This allows us to estimate a time scale for the validity of the ``pure pulse" approximation discussed in the previous section, given by $t_{{\rm valid}} \sim T/(U_0 \, t_p)$. 

We now validate the effective Hamiltonian in Eq.~\eqref{eq:Heffimpurepulse} using exact diagonalization. As in the main text, we study the dynamics of two fermions on a single plaquette (four lattice sites), and we calculate the mean probability of changing parity over time. Figure~\ref{fig: realistic pulse} compares the exact dynamics generated by the full time-dependent Hamiltonian in Eq.~\eqref{eq: exact Ham impure pulse} with the dynamics generated by the effective Hamiltonian in Eq.~\eqref{eq:Heffimpurepulse}, for two values of the ratio $t_p/T$. The good agreement between these dynamics confirms that the $Z_2$-breaking effects associated with the finite duration and impure nature of the pulses are well captured by the effective Hamiltonian in Eq.~\eqref{eq:Heffimpurepulse}.

We conclude this Section by proposing realistic experimental parameters for an optical-lattice implementation. First, we set a reasonable value for the maximal inter-chain hopping strength, $J_{\rm{II}}\!=\!3 \pi/2t_p\!\sim \!$ kHz ($\hbar\!=\!1$). This imposes a constraint on the pulse duration, $t_p\!\sim\!1 \, \rm{ms}$. The ``pure pulse" approximation would then require a period of the order $T\!\sim\!10\, \rm{ms}$. Besides, the validity of the high-frequency approximation would require a bare interaction of order $U_0\!\sim\!10\, \rm{Hz}$. 

Finally, we note that the ``pure pulse" sequence discussed in the previous Section could also be implemented by simply deactivating $\hat H_0$ during each pulse, i.e. through a proper time-modulation of the bare interaction strength $U_0 (t)$; see also Appendix~\ref{app_pair}.

\end{document}